\documentclass[11pt]{article}
\usepackage{amsmath}
\usepackage{amssymb}
\usepackage{amsfonts}
\usepackage{amscd}
\usepackage{accents}
\usepackage{pst-all}
\usepackage{epsfig}
\usepackage{graphicx}
\usepackage[tight]{subfigure}
\date{\today}
\newcommand{\bmat}{\left(\begin{array}}
\newcommand{\emat}{\end{array}\right)}
\newcommand{\be}{\begin{equation}}
\newcommand{\ee}{\end{equation}}
\newcommand{\ba}{\begin{eqnarray}}
\newcommand{\ea}{\end{eqnarray}}

\def\lsim{\raise0.3ex\hbox{$\;<$\kern-0.75em\raise-1.1ex\hbox{$\sim\;$}}}
\def\gsim{\raise0.3ex\hbox{$\;>$\kern-0.75em\raise-1.1ex\hbox{$\sim\;$}}}

\def\be{\beta}

\begin{document}

\vspace*{-.6in} \thispagestyle{empty}
\begin{flushright}
DESY 13-090
\end{flushright}
\begin{flushright}
LPT 13-38
\end{flushright}
\baselineskip = 20pt

\vspace{.5in} {\LARGE
\begin{center}
{\bf  More on the Hypercharge Portal into the Dark Sector  }
\end{center}}

\vspace{.5in}

\begin{center}
{\bf  Florian Domingo$^{~a}$, Oleg Lebedev$^{~a}$, Yann Mambrini$^{~b}$,
J\'er\'emie Quevillon$^{~b}$   and   
Andreas Ringwald$^{~a}$   }  \\

\vspace{.5in}

a: \emph{DESY Theory Group, 
Notkestrasse 85, D-22607 Hamburg, Germany
 }
\\ 
b: \emph{Laboratoire de Physique Th\'eorique, 
Universit\'e Paris-Sud, F-91405 Orsay, France  }
\end{center}

\vspace{.5in}

\begin{abstract}
\noindent
If the hidden sector contains more than one U(1) groups, additional dim-4 couplings
(beyond the kinetic mixing)
between the massive U(1) fields and the hypercharge generally appear. 
These are of the form
similar to the Chern--Simons interactions. We study the phenomenology
of such couplings including constraints from laboratory experiments and implications
for dark matter. The hidden vector fields can play the role of dark matter  whose
characteristic signature would be monochromatic gamma ray emission from the 
galactic center. We show that this possibility is consistent with the LHC and other
laboratory constraints, as well as astrophysical bounds. 
\end{abstract}

\newpage

\section{Introduction}

The existence of new physics structures beyond those of the Standard Model (SM) is motivated, among 
other things, by the puzzles of dark matter (DM) and inflation.  The minimal way to address these
problems is to add a ``hidden'' sector containing the required SM--singlet fields. 
The existence of the hidden sector can also be motivated from the top--down viewpoint, in particular,
by realistic string constructions \cite{Gross:1984dd,Buchmuller:2005jr}. 
Such a sector can  couple to the SM fields through  products of gauge--singlet operators,
including those of dimension 2 and 3. In this work, we study in detail the corresponding  
couplings to the hypercharge field.

Let us define the ``hidden sector'' as a set of fields which carry no SM gauge quantum numbers. Then 
a ``portal'' \cite{Patt:2006fw}  would be an operator that couples the SM fields to such SM singlets.  
Let us consider  the minimal case: 
suppose  that the relevant low energy degrees of freedom in the hidden sector are those of a
Weyl fermion $\chi$, or a massive vector $V_\mu$, or a real scalar $S$ (one field at a time).  
Then the lowest, up to $dim$--4, dimension operators which couple  the SM to the hidden sector 
are given by
\begin{eqnarray} 
&& O_1 = \Psi_L H \chi + {\rm h.c.} \;, \nonumber\\
&& O_2 = F_{\mu\nu}^Y \; F^{V \; \mu\nu} \;, \nonumber\\
&& O_3 = {\overline \Psi}_i \gamma_\mu (1 +  \alpha_{ij} \gamma_5) \Psi_j \; V^\mu  + {\rm h.c.}    \;, \nonumber\\
&& O_4 = H^\dagger H \; V_\mu V^\mu + \beta\; H^\dagger i D_\mu H \; V^\mu + {\rm h.c.} \;, \nonumber\\
&& O_5 = H^\dagger H \; S^2 +  \mu_S\; H^\dagger H \; S \;. 
\end{eqnarray}
Here $\Psi_L$ is the lepton doublet;  
$F^{Y}_{\mu\nu}$ and 
$F^{V}_{\mu\nu}$ are  the field strength tensors for   hypercharge and $V_\mu$, respectively;
$\Psi_i$ is an SM fermion with generation index $i$;  $D_\mu$ is the covariant derivative with respect to 
the SM gauge symmetries, and $\alpha_{ij}, \beta, \mu_S$ are constants.
Note that a particular version of operator $O_3$ is induced by $O_2$ after diagonalization of the
vector kinetic terms.

An attractive feature of such an extension of the Standard Model is that it can offer viable 
dark matter candidates as well as provide a link to the inflaton sector.
In particular, a sufficiently light ``right--handed neutrino'' $\chi$  is long--lived
and can constitute warm dark matter \cite{Asaka:2005an}. 
Also, a massive vector $V_\mu$ (or a scalar $S$ \cite{Silveira:1985rk}) can
inherit a $Z_2$ symmetry from hidden sector gauge interactions, which would eliminate terms linear
in $V_\mu$ and make it a stable cold dark matter candidate \cite{Lebedev:2011iq}. Finally, 
the Higgs coupling $H^\dagger H \; S^2$ to the inflaton $S$ would be instrumental 
in reconciling metastability of the electroweak vacuum with inflation \cite{Lebedev:2012sy}.

In this work, we explore a more general $dim$--4 hypercharge coupling to the hidden sector,
when the latter contains multiple U(1)'s. In this case, a Chern--Simons--type coupling becomes
possible \cite{Coriano':2005js,Anastasopoulos:2006cz,Antoniadis:2009ze,Dudas:2009uq,Antoniadis:2010zzf}. 
If such a coupling is the only SM portal into the hidden sector, the lightest U(1) vector
field can play the role of dark matter. The trademark signature of this scenario is the presence 
of monochromatic gamma--ray lines in the photon spectrum of  the galactic center. We analyze 
general experimental constraints on the Chern--Simons--type coupling as well as the constraints
applicable when the vector field constitutes dark matter.

\section{Hypercharge couplings to the ``hidden'' sector}

Suppose the ``hidden'' sector contains two massive U(1) gauge fields $C_\mu$ and $D_\mu$. 
Before electroweak symmetry breaking, the most general dim-4 interactions of these fields with the hypercharge 
boson
$B_\mu$ are described by the Lagrangian
\begin{eqnarray}
 {\cal L}&=& -{1\over 4} B_{\mu \nu} B^{\mu \nu} -{1\over 4} C_{\mu \nu} C^{\mu \nu}
 -{1\over 4} D_{\mu \nu} D^{\mu \nu}   
-{\delta_1\over 2} B_{\mu \nu} C^{\mu \nu}  -{\delta_2\over 2} B_{\mu \nu} D^{\mu \nu}
-  {\delta_3\over 2} C_{\mu \nu} D^{\mu \nu} \nonumber\\
& + & {M_C^2\over 2} C_\mu C^\mu  +  {M_D^2\over 2} D_\mu D^\mu + \delta M^2 C_\mu D^\mu 
+ \kappa\; \epsilon_{\mu\nu \rho\sigma} B^{\mu\nu} C^\rho D^\sigma \;.  
\label{L0}
\end{eqnarray}
Here we have assumed CP symmetry such that terms of the type $B^{\mu\nu} C_\mu D_\nu$ are 
not allowed (see \cite{Farzan:2012kk} for a study of the latter). The kinetic and mass mixing can be eliminated by field redefinition \cite{Holdom:1985ag},
which to first order in the mixing parameters $\delta_i$ and $\delta M^2$ reads
\begin{eqnarray}
&& B_\mu \rightarrow B_\mu + \delta_1 \; C_\mu + \delta_2 \; D_\mu \;, \nonumber \\
&& C_\mu  \rightarrow C_\mu + { \delta_3 \; M_D^2 - \delta M^2 \over M_D^2-M_C^2  }~ D_\mu \;, \nonumber \\
&& D_\mu  \rightarrow D_\mu - { \delta_3 \; M_C^2 - \delta M^2 \over M_D^2-M_C^2  }~ C_\mu \;.
\end{eqnarray}
In terms of the new fields, the Lagrangian  is
\begin{equation}
 {\cal L}= -{1\over 4} B_{\mu \nu} B^{\mu \nu} -{1\over 4} C_{\mu \nu} C^{\mu \nu}
 -{1\over 4} D_{\mu \nu} D^{\mu \nu}   +
 {M_C^2\over 2} C_\mu C^\mu  +  {M_D^2\over 2} D_\mu D^\mu 
+ \kappa\; \epsilon_{\mu\nu \rho\sigma} B^{\mu\nu} C^\rho D^\sigma,  \label{L}
\end{equation}
which will be the starting point for our phenomenological analysis. We note that, due to the kinetic
mixing $\delta_{1,2}$,  $C_\mu$ and $D_\mu$ have small couplings to the Standard Model matter.
Since we are mainly interested in the effect of the Chern--Simons--type   term 
$ \epsilon_{\mu\nu \rho\sigma} B^{\mu\nu} C^\rho D^\sigma $, we will set $\delta_{1,2}$ to be very small
or zero in most of our analysis.

The term $ \epsilon_{\mu\nu \rho\sigma} B^{\mu\nu} C^\rho D^\sigma $ has  dimension 4. However, it 
vanishes in the limit of zero vector boson masses 
by gauge invariance, both for the Higgs and St\"uckelberg mechanisms.
This means  that it comes effectively
from  a  higher dimensional  operator with $\kappa$ proportional to 
$M_C M_D/\Lambda^2$, where $\Lambda$ is the cutoff
 scale or the mass scale of heavy particles we have integrated out. 
On one hand, this operator does not decouple as $\Lambda \rightarrow \infty$
since both $M_{C,D}$ and $\Lambda$ are given by the ``hidden'' Higgs VEV times
the appropriate couplings; on the other hand,  
$ \epsilon_{\mu\nu \rho\sigma} B^{\mu\nu} C^\rho D^\sigma $ is phenomenologically relevant only  
if $M_{C,D}$ are not far above the weak scale. Thus, 
this term  represents a meaningful approximation   in a particular energy window, which we will quantify later. 
(A similar situation occurs in the vector Higgs portal models, where the interaction 
$H^\dagger H V_\mu V^\mu$ has naive dimension 4, but originates from 
a  dim-6 operator \cite{Lebedev:2011iq}.)
From the phenomenological perspective, it is important that  
$ \epsilon_{\mu\nu \rho\sigma} B^{\mu\nu} C^\rho D^\sigma $ is  the leading operator at 
low energies, e.g. relevant to  non--relativistic annihilation of dark matter composed of $C_\mu$ or $D_\mu$,
and thus we will restrict our attention to this coupling only.

A coupling of this sort appears in various models upon integrating
out  heavy fields charged under both U(1)'s and hypercharge. Explicit anomaly--free examples
can be found in  \cite{Dudas:2009uq} and \cite{Antoniadis:2009ze}. In these cases, the 
Chern--Simons term arises upon integrating out heavy, vector--like with respect to the SM,
fermions. Both the vectors and the fermions get their masses from the Higgs mechanism,
while the latter can be made heavy by choosing large Yukawa couplings compared to the gauge
couplings. In this limit, Eq.~$\ref{L}$ gives the corresponding low energy 
action.\footnote{We note that certain ``genuine'' gauge invariant dim-6 operators such as 
${1\over \Lambda^2} \epsilon^{\mu\nu\rho\sigma} B_{\mu\nu} C_{~\rho}^{\tau} D_{\tau\sigma }$
reduce to the Chern-Simons term on--shell in the non--relativistic limit 
($C_{\mu\nu} \rightarrow C_{0i}=iM_C C_i~$; $C_0=0$ and similarly for $D_{\mu\nu}$). Such operators
should generally be taken into account when deriving the low energy action in explicit 
microscopic models.}

Finally, we note that increasing the number of hidden U(1)'s does not bring in  hypercharge--portal 
interactions  with a new structure, so our considerations apply quite generally.

\section{Phenomenological constraints}  

In this section we derive constraints on the coupling constant $\kappa$ from various 
laboratory experiments as well as unitarity considerations.
The relevant interaction to leading order is  given by 
\begin{equation} 
\Delta {\cal L}= 
\kappa\; \cos\theta_W\;  
\epsilon_{\mu\nu \rho\sigma} F^{\mu\nu} C^\rho D^\sigma -
\kappa\; \sin\theta_W\;  
\epsilon_{\mu\nu \rho\sigma} Z^{\mu\nu} C^\rho D^\sigma \;,
\label{coupling}
\end{equation}
where $ F^{\mu\nu}$ and $Z^{\mu\nu} $ are the photon and Z-boson field 
strengths, respectively.

In what follows, we set the kinetic mixing to be negligibly small such that the lighter of
the $C$ and $D$ states is not detected and thus appears as missing energy and momentum.
There are then two possibilities: the heavier state decays into the lighter state plus $\gamma$
either outside or inside the detector.
Consider first the case where
 the mass splitting and  $\kappa$ are  relatively small such that
both $C$ and $D$ are ``invisible''.

\subsection{Unitarity}

The coupling $ \epsilon_{\mu\nu \rho\sigma} B^{\mu\nu} C^\rho D^\sigma $ involves longitudinal
components of the massive vectors. Therefore, some scattering amplitudes will grow indefinitely
with energy, which imposes a cutoff on our effective theory. For a fixed cutoff, this translates
into a bound on $\kappa$.

Consider the scattering process
\begin{equation}
C_\mu ~ C_\nu \rightarrow D_\rho ~ D_\sigma
\end{equation}
at high energies, $E \gg M_{C,D}$.  The vertex can contain  longitudinal components of at most 
one vector since $\epsilon_{\mu\nu\rho\sigma} (p_1+p_2)^\mu p_1^\nu p_2^\rho =0$. Then one
finds that the amplitude grows quadratically with energy,
\begin{equation}
{\cal A} \sim \kappa^2 ~{E^2\over M_{C,D}^2} \;,
\end{equation}
with the subscripts $C$ and $D$ applying to the processes involving longitudinal components
of $C_\mu$ and $D_\mu$, respectively. On the other hand, the amplitude cannot
exceed roughly $8 \pi$.
Neglecting order one factors, the resulting  constraint is
\begin{equation} 
{\kappa  \over M} < { \sqrt{8 \pi}\over \Lambda  } \;, 
\label{uni}
\end{equation}
where $M=\min \{ M_C , M_D \}$ and $\Lambda$ is the cutoff scale. As explained in the previous 
section, $\Lambda$ is associated with the mass scale of new states charged under $\rm{U(1)}_Y$.
Since constraints on such states are rather stringent, it is reasonable to take
$\Lambda \sim 1$ TeV. This implies that light vector bosons can couple only very weakly,
e.g. $\kappa < 10^{-5}$ for $M \sim 1$ MeV.

It is important to note that the unitarity bound applies irrespective of
whether $C$ and $D$ are stable or not. Thus it  applies to the case
$M_D \gg M_C$ or vice versa and  also in the presence of the kinetic mixing.

\subsection{Invisible $\Upsilon$ decay}

Suppose that $D$ is the heavier state and
the decay $D \rightarrow C +\gamma$  is not fast enough
to occur inside the detector. Then production of $C$ and $D$ would appear as missing energy.
In particular, light $C,D$ can be produced in the invisible $\Upsilon$ decay
\begin{equation} 
\Upsilon \rightarrow {\rm inv} ~~,
\end{equation}
which is a powerful probe of new physics since its branching ratio 
in the Standard Model is small, about $10^{-5}$  \cite{Chang:1997tq}.
In our case,
this decay is dominated by the $s$--channel   annihilation through the photon,
while the $Z$--contribution is suppressed by $m_\Upsilon^4/m_Z^4$.
We find
\begin{eqnarray}
\Gamma (\Upsilon \rightarrow CD) &=& 2\alpha \kappa^2 \cos^2 \theta_W \;Q_d^2 \; {f_\Upsilon^2 \over m_\Upsilon}~
\sqrt{ 1- 2 {M_C^2 + M_D^2 \over m_\Upsilon^2} + {(M_C^2-M_D^2)^2\over m_\Upsilon^4 }}
\nonumber \\
&\times& \left[ 1 + {m_\Upsilon^2 \over 12}~ \left( {1\over M_C^2} + {1\over M_D^2}  \right)
\left(  1- 2 {M_C^2 + M_D^2 \over m_\Upsilon^2} + {(M_C^2-M_D^2)^2\over m_\Upsilon^4 }    
\right) \right] \;,
\end{eqnarray}
where $\alpha$ is the fine structure constant, $Q_d$ is the down quark charge and $f_\Upsilon$
is the $\Upsilon$ decay constant, $ \langle 0 \vert \bar b \gamma^\mu b
\vert \Upsilon  \rangle  = f_\Upsilon m_\Upsilon \epsilon^\mu$ with 
$\epsilon^\mu$ being the $\Upsilon$ polarization vector.
In the limit $M_{C,D}^2 \ll m_\Upsilon^2$ and $M_C \simeq M_D = M$, the 
decay rate becomes 
\begin{equation}
\Gamma (\Upsilon \rightarrow CD) \simeq {1\over 3} \alpha\kappa^2
\cos^2 \theta_W \;Q_d^2 \; { f_\Upsilon^2 m_\Upsilon \over M^2  }  \;.
\end{equation}
Taking $m_\Upsilon(1S)=9.5$ GeV, $\Gamma_\Upsilon(1S)=5.4\times 10^{-5}$ GeV,
 $f_\Upsilon=0.7$ GeV and using the BaBar limit 
BR$(\Upsilon \rightarrow {\rm inv}) <3\times 10^{-4}$ at 90\% CL 
\cite{Aubert:2009ae},
we find
\begin{equation}
{\kappa \over M} < 4 \times 10^{-3}~ {\rm GeV^{-1}} \;.
\end{equation} 
This bound applies to vector boson masses up to a few GeV and 
disappears above $m_\Upsilon /2$.
An analogous bound from $J/\Psi  \rightarrow {\rm inv} $ is weaker. 

We note that the $\Gamma \propto 1/M^2$ dependence is characteristic to
production of the longitudinal components of massive vector bosons. 
The corresponding polarization vector grows with energy as $E/M$, or
in other words, at  $M \ll m_\Upsilon$, 
the decay is dominated by the Goldstone boson production,
whose couplings grow with energy. Thus, stronger constraints on $\kappa$ 
are expected from the decay of  heavier states.

The corresponding bound from the radiative $\Upsilon$ decay $\Upsilon
\rightarrow \gamma + {\rm inv}$ is much weaker.  By $C$--parity, such a decay
can only be mediated by the $Z$ boson, which brings in the 
$m_\Upsilon^4/m_Z^4$ suppression factor. The resulting constraint is 
negligible.

\subsection{ Invisible $Z$ decay}

The invisible width of the $Z$ boson $\Gamma^Z_{\rm inv}$   is strongly constrained by the LEP measurements \cite{ALEPH:2005ab}. The process $Z \rightarrow C D $ contributes 
to $\Gamma^Z_{\rm inv}$ for vector boson masses up to about 45 GeV, thereby leading to a 
bound on $\kappa$. We find
\begin{eqnarray}
\Gamma (Z \rightarrow CD) &=& {1\over 2 \pi} \; \kappa^2 \sin^2 \theta_W \; m_Z                 ~
\sqrt{ 1- 2 {M_C^2 + M_D^2 \over m_Z^2} + {(M_C^2-M_D^2)^2\over m_Z^4 }}
\nonumber \\
&\times& \left[ 1 + {m_Z^2 \over 12}~ \left( {1\over M_C^2} + {1\over M_D^2}  \right)
\left(  1- 2 {M_C^2 + M_D^2 \over m_Z^2} + {(M_C^2-M_D^2)^2\over m_Z^4 }    
\right) \right] \;.
\end{eqnarray}
In the limit $M_{C,D}^2 \ll m_Z^2$ and $M_C \simeq M_D = M$, it becomes
\begin{equation}
\Gamma (Z \rightarrow CD) \simeq {\kappa^2 \sin^2 \theta_W \over 12 \pi }~ {m_Z^3 \over M^2} \;.
\end{equation}
Taking the bound on the BSM contribution to  $\Gamma^Z_{\rm inv}$
to be roughly 3 MeV (twice the experimental error--bar of  $\Gamma^Z_{\rm inv}$ \cite{ALEPH:2005ab}), we have
\begin{equation}
{\kappa \over M} <  8\times 10^{-4}~ {\rm GeV^{-1}} \;.
\end{equation}
In the given kinematic range,
this constraint  is even stronger than the unitarity bound for $\Lambda=1$ TeV
and comparable to the latter with a multi--TeV cutoff. As explained above,
such sensitivity of $ Z \rightarrow {\rm inv} $ to $\kappa$
 is due to the $E/M$ enhancement of  the longitudinal vector boson production.

\subsection{ $B \rightarrow K + {\rm inv}$ and $K \rightarrow \pi + {\rm inv}$  }

Flavor changing transitions with missing energy are also a sensitive probe of 
matter couplings to ``invisible'' states (see e.g. \cite{Andreas:2010ms}). 
The decay $B \rightarrow K + C\;D $ proceeds via the SM flavor violating
$\bar b s Z $ and $\bar b s \gamma $ vertices with subsequent conversion of $Z,\gamma$
into $C$ and $D$. Numerically, the process is dominated by the $Z$ contribution
with the flavor changing vertex \cite{Inami:1980fz,Grinstein:1988me}
\begin{equation}
{\cal L}_{\bar b s Z}= \lambda_{\bar b s Z}
~ \bar b_L \gamma_\mu s_L~Z^\mu \;,
\end{equation}
with
\begin{equation}
\lambda_{\bar b s Z} = {g^3 \over 16 \pi^2 \cos\theta_W}\; V^*_{tb} V_{ts} ~f\left({m_t^2\over m_W^2}\right) \;,
\end{equation}
where $V_{ij}$ are the CKM matrix elements and $f(x)$ is the Inami--Lim function 
\cite{Inami:1980fz},
\begin{equation}
f(x) = {x\over 4} \; \left(   {x-6 \over x-1}  + {3x+2 \over (x-1)^2} \; \ln x             \right) \;.
\end{equation}
We find 
\begin{eqnarray}
\Gamma (B \rightarrow K + C\;D) &=&   {\kappa^2 \lambda_{\bar b s Z}^2 \sin^2\theta_W
\over 2^7 \pi^3 m_B^3 m_Z^4 }  \int^{(m_B-m_K)^2}_{(M_C+M_D)^2} {ds \over  s}~ f_+^2(s) \\
&\times& \sqrt{ (s-M_C^2-M_D^2)^2-4M_C^2 M_D^2}
~ \Bigl( (s+m_B^2-m_K^2)^2-4m_B^2s  \Bigr)^{3/2} \nonumber\\
& \times& \biggl[ 1+ {1\over 12 s} \left( {1\over M_C^2} + {1\over M_D^2}  \right) 
\Bigl( (s-M_C^2-M_D^2)^2-4M_C^2 M_D^2 \Bigr) \biggr] \;, \nonumber
\end{eqnarray}
where the form factor $f_+(s)$ is defined by 
$\langle K(p_K) \vert \bar b \gamma^\mu s \vert B(p_B)  \rangle  = (p_K+p_B)^\mu f_+(s)+
(p_B-p_K)^\mu f_-(s)$ with $s=(p_B-p_K)^2$.
The decay rate is dominated by the contribution from large invariant masses of the $C,D$ pair
due to the longitudinal vector boson production. This justifies the subleading character of the 
photon contribution: the corresponding dipole operator  can be significant at low invariant
masses due to the $1/s$ pole, as in the $B\rightarrow K l^+ l^-$ processes (see e.g. 
\cite{Buchalla:2010jv} for a recent summary). The relative size of various $\Delta F=1$ 
operators can be found in \cite{Grinstein:1988me,Inami:1980fz},  and we find 
that the photon contribution is unimportant. 

The relevant experimental limit has been obtained by BaBar: 
BR$(B^+ \rightarrow K^+ \nu \bar \nu)  < 1.3 \times 10^{-5}$ at 90\% CL 
\cite{delAmoSanchez:2010bk}. Then taking $f_+(0)=0.3$ and using its $s$--dependence
from \cite{Buchalla:2010jv}, we find
\begin{equation}
{\kappa \over M} <  1~ {\rm GeV^{-1}} \;,
\end{equation}
for $M_C\simeq M_D =M$ up to roughly 2 GeV.
The above considerations equally apply to the process $K\rightarrow \pi + {\rm inv} $,
up to trivial substitutions. We find that the resulting bound is weak, 
$ \kappa /M < 30 ~ {\rm GeV^{-1}} $. This stems from the $m_{\rm meson}^7/(M^2 m_Z^4)$ 
behavior of the rate, which favors heavier mesons.

Finally, the Chern--Simons coupling does not contribute to $B \rightarrow CD$ due to the 
$\epsilon$--tensor contraction, so there is no bound from the $ B \rightarrow {\rm inv} $ 
decay. Also, $\kappa$ contributes to $(g-2)_\mu$ only at the two loop level such that 
 the resulting bound is insignificant.

\begin{figure}
    \centering
    \includegraphics[height=22em]{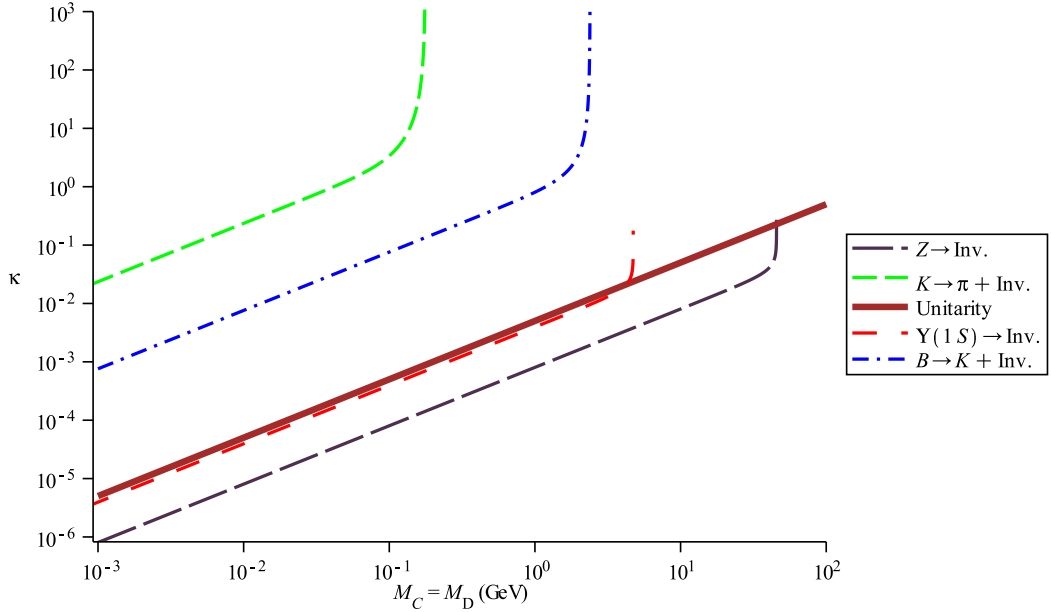}
  \caption{ Bounds on $\kappa$. The unitarity bound assumes $\Lambda=1$ TeV.}
  \label{kappabounds}
\end{figure}

The summary of the bounds is shown in Fig.~\ref{kappabounds}. We see that the most stringent
limits are set by the $Z$ invisible width and unitarity considerations. The latter 
has the advantage of not being limited by kinematics and places a tight bound on $\kappa$
for vector masses up to  about 100 GeV.

\subsection{Bounds on decaying vector bosons $D \rightarrow C + \gamma$}

When the  vector boson   mass difference is not too small, the heavier particle, say $D$, will 
decay inside the detector. In this case, the constraints on $\kappa$ get somewhat modified.
The decay width $\Gamma_D$   is given by
\begin{equation}
 \Gamma (D \rightarrow C + \gamma) = { \kappa^2 \cos^2\theta_W \over 24 \pi}~ 
{ (M_D^2-M_C^2)^3 \over M_D^3 }~ \left(  {1\over M_C^2} + {1\over M_D^2}   \right) \;,
\end{equation}
assuming that the $Z$--emission is kinematically forbidden.
Given the velocity $v_D$ and lifetime $\tau_D$,
$D$ decays inside the detector if $v_D \tau_D = \vert {\bf p}_D \vert /(M_D \Gamma_D )$
is less than the detector size $l_0$, which we take to be $\sim 3$ m.
In this case, $\kappa$ is constrained by radiative decays with missing energy.

Consider the radiative  decay $\Upsilon(1S) \rightarrow \gamma + {\rm inv}$.
Its branching ratio is constrained by BaBar: BR$(\Upsilon(1S) \rightarrow \gamma + {\rm inv})
< 6 \times 10^{-6}$ for a 3--body final state and $M_C$ up to about 3 GeV 
\cite{delAmoSanchez:2010ac}. Since BR$(D \rightarrow C + \gamma) \sim $ 100\%, this requires
approximately
\begin{equation}
{\kappa \over M} < 6 \times 10^{-4}~ {\rm GeV^{-1}} \;,
\label{ups-rad}
\end{equation} 
which is the strongest bound  on $\kappa$ in the kinematic range $M\lsim 3$ GeV.
This bound applies for
\begin{equation}
\Delta M \gsim  \left( 3 \pi m_\Upsilon M \over 4 \kappa^2 \cos^2\theta_W \; l_0   \right)^{1/3} \;,
\end{equation}
where we have made the approximation $M_D-M_C = \Delta M \ll M \ll m_\Upsilon$. 
For example, taking the maximal allowed $\kappa$ consistent with 
(\ref{ups-rad}) at $M=1$ GeV, the decay occurs within the detector for
$\Delta M > 2$ MeV. (However, since the experimental  cut on the photon energy
is 150 MeV, $\Delta M$ close to this bound would not lead to a detectable 
signal.)

On the other hand,
the bound on $\kappa$ from the invisible $Z$ width does not change even for decaying $D$.
The reason is that the invisible width is defined by subtracting  the visible decay width 
into fermions 
$\Gamma ( Z \rightarrow  \bar f f)$ from the total width $\Gamma_Z$   measured via the 
energy dependence of the hadronic cross section \cite{ALEPH:2005ab}. 
Thus, $Z \rightarrow \gamma + 
{\rm inv}$ qualifies as ``invisible'' decay and we still have
   \begin{equation}
{\kappa \over M} <  8\times 10^{-4}~ {\rm GeV^{-1}} \;,
\end{equation}
as long as the decay is kinematically allowed.

Finally, the unitarity bound 
\begin{equation} 
{\kappa  \over M} < { \sqrt{8 \pi}\over \Lambda  } 
\end{equation}
remains intact as well. Another constraint in the higher mass range 
$m_Z/2 \lsim  M \lsim 100$ GeV is imposed by the LEP monophoton 
searches $e^+ e^- \rightarrow \gamma + {\rm inv}$ \cite{Achard:2003tx}.
We find, however, that it is somewhat weaker than the unitarity bound for 
$\Lambda=1$ TeV (the same applies to $e^+ e^- \rightarrow {\rm inv}$).

Thus, the strongest constraints in Fig.~\ref{kappabounds} apply also to the case
of decaying vector bosons, while the $\Upsilon$ bound becomes competitive and even
the tightest one at lower masses.
For $M \gsim 100$ GeV,
some of the  relevant LHC constraints will be discussed in the next section, while
their comprehensive analysis requires  a separate study.

Let us conclude by remarking on the astrophysical constraints. These apply to very 
light, up to ${\cal O}$(MeV), particles.  In particular, the rate of energy loss in
horizontal--branch stars sets stringent bounds on light particle emission in
Compton--like scattering $\gamma + e \rightarrow e + C +D$. We find that this cross
section in the non--relativistic limit scales approximately as 
$ \alpha^2 \kappa^2/(6 \pi m_e^2) \; (T/M)^{2} $, with $T \sim \;$keV being the core temperature. 
Comparison to the axion models  \cite{Raffelt:1994ry} leads   
then to the bound $\kappa/M < 10^{-7}\;
 {\rm GeV}^{-1}  $ for $M \ll \;$keV, which is much stronger than the 
laboratory  constraints in this mass range. Analogous supernova cooling considerations 
extend the  range to  ${\cal O}$(MeV). A dedicated study of astrophysical constraints will
be presented elsewhere.

\section{Vector Dark Matter and the Chern--Simons coupling}

In this section,  we consider a special case of the Lagrangian (\ref{L0}) 
with
\begin{equation}
 \delta_{1,2}=0 \;,
\end{equation}
that is, the new gauge bosons do not mix with the hypercharge. This can be enforced
by the $Z_2$ symmetry
\begin{equation}
C_\mu \rightarrow - C_\mu ~~~,~~~ D_\mu \rightarrow - D_\mu  \;.
\end{equation}
It is straightforward to construct  microscopic models  which lead to an effective
theory endowed with this symmetry at one loop. However, to make the $Z_2$ persist at
higher loop levels is much more challenging and beyond the scope of this paper. 

\begin{figure}
    \centering
    \includegraphics[scale=0.5]{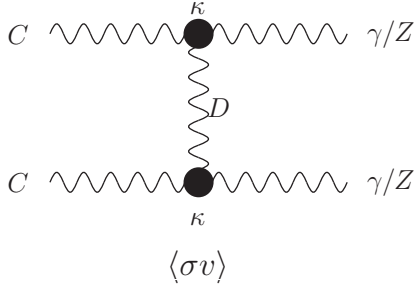}
  \caption{Dark matter annihilation into photons and $Z$--bosons.}
  \label{Fig:feynman}
\end{figure}

The relevant Lagrangian in terms of the propagation eigenstates is again given by (\ref{L}),
except now $C$ and $D$ do not couple to ordinary matter. The $Z_2$ symmetry forbids their kinetic 
mixing with the photon and the $Z$. This makes  the lighter state, $C$, stable and a good 
dark matter candidate. In what follows, we consider $M_C$ of order the electroweak scale such
that dark matter is of WIMP type.

Our vector dark matter interacts with the SM only via the Chern--Simons type terms (\ref{coupling}).
These allow for DM annihilation into photons and $Z$ bosons (Fig.~\ref{Fig:feynman} and 
its cross--version). The corresponding 
cross sections for $M_C \simeq M_D=M$ in the non--relativistic limit are given 
by\footnote{For simplicity, we have assumed a single mass scale for the vectors with $D$ being somewhat 
heavier such that it decays into $C$ and a photon. Further  details are unimportant for our purposes.}
\begin{align}
 &\langle \sigma v\rangle (CC\to\gamma\gamma)\simeq\frac{29\kappa^4 \cos^4\theta_W }{36\pi M^2} \;, 
\label{annih}\\
 &\langle \sigma v \rangle(CC\to\gamma Z)\simeq\frac{\kappa^4\sin^2\theta_W \; \cos^2\theta_W}{18\pi M^2}\left(1-\frac{M_Z^2}{4M^2}\right)\left[29-\frac{5M_Z^2}{2M^2}+\frac{5M_Z^4}{16M^4}\right]\Theta(2M-M_Z) \;, \nonumber\\
 &\langle \sigma v\rangle (CC\to ZZ)\simeq\frac{\kappa^4 \sin^4\theta_W }{36\pi M^2}\sqrt{1-\frac{M_Z^2}{M^2}}\left(1-\frac{M_Z^2}{2M^2}\right)^{-2}\left[29-34\frac{M_Z^2}{M^2}+14\frac{M_Z^4}{M^4}\right]\Theta(M-M_Z) , \nonumber
\end{align}
where $\Theta$ is the Heaviside distribution.
These processes both regulate dark matter abundance and lead to potentially observable
gamma--ray signatures, which we study in detail below. 

The distinctive
feature of the model is the presence of monochromatic gamma--ray  lines in the spectrum
of photons coming from the Galactic Center (see e.g.~\cite{Vertongen:2011mu}). 
In particular, for heavy dark matter ($M^2 \gg M_Z^2$),
the final states $\gamma \gamma$, $\gamma Z$ and $ZZ$ are produced in the proportion
$\cos^4\theta_W$,  $2\sin^2\theta_W \; \cos^2\theta_W$ and $\sin^4\theta_W$, respectively.
This implies that  continuous gamma--ray emission is subdominant and constitutes about a third 
of the annihilation cross section, while the monochromatic gamma--ray emission dominates.

\subsection{WMAP/PLANCK constraints}

\begin{figure}
    \centering
    \includegraphics[scale=0.6]{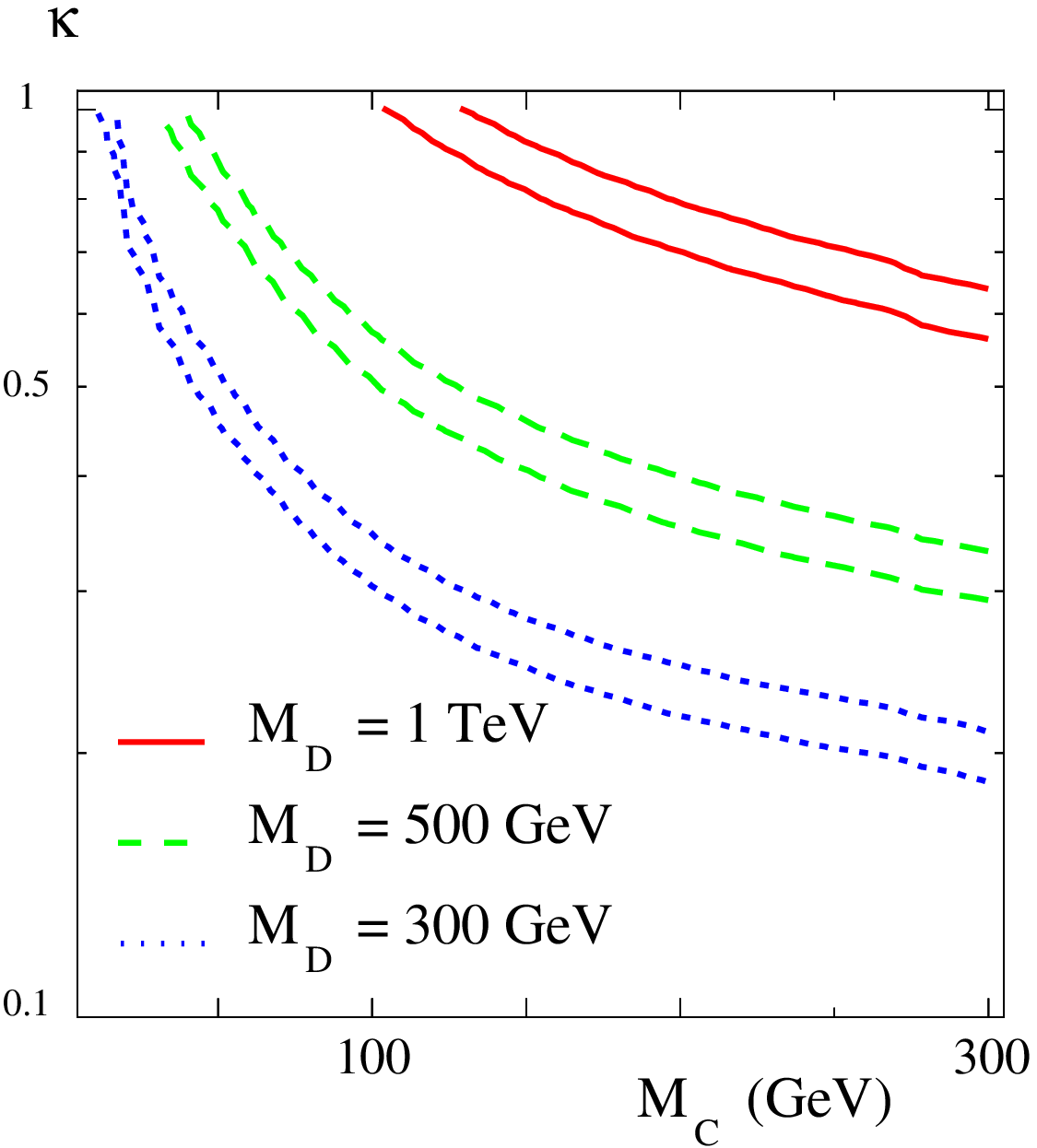}
    \hspace{0.5cm}
      \includegraphics[scale=0.6]{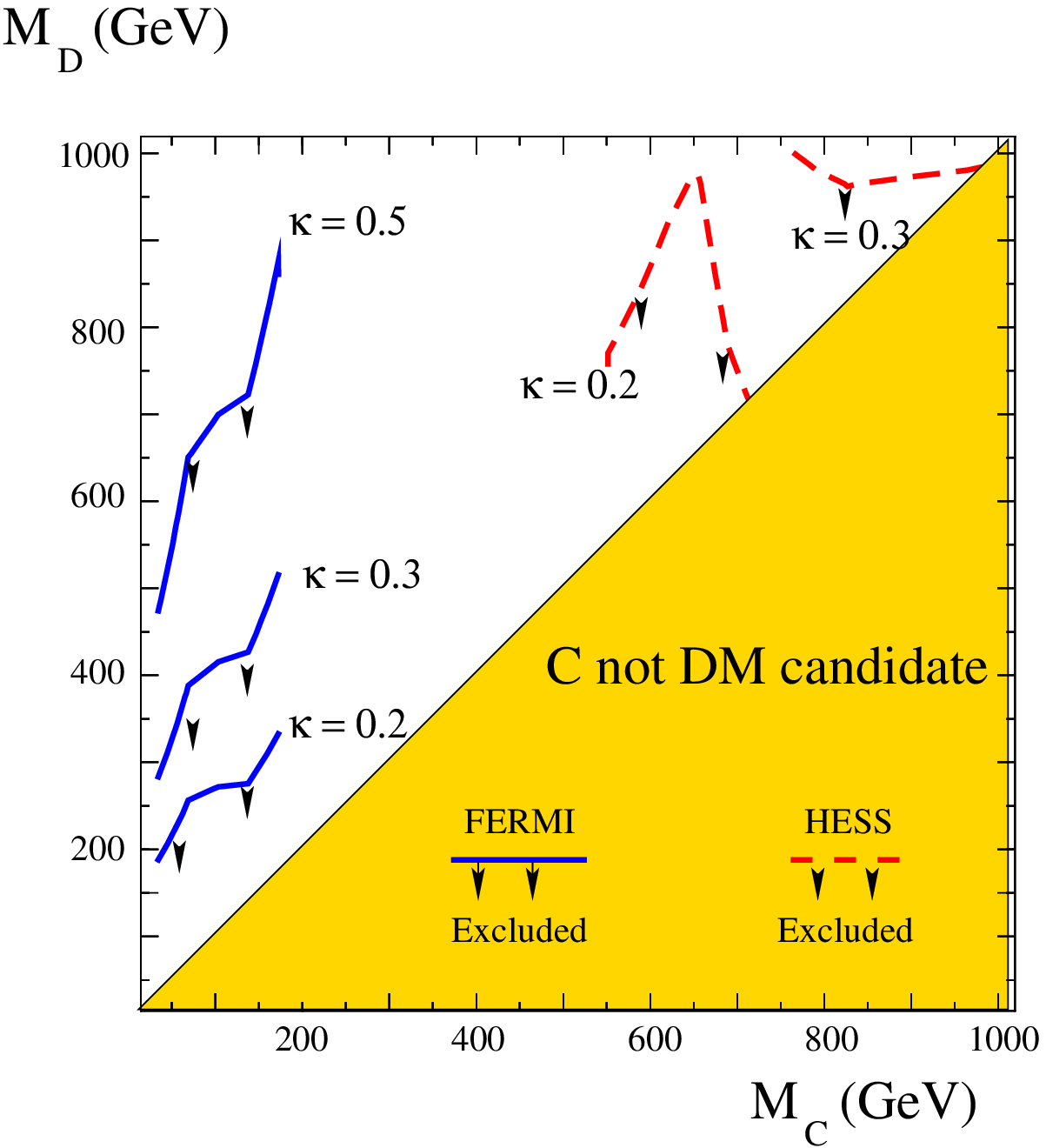}
  \caption{Left: the areas between the lines represent 
values of $\kappa$ consistent with the  WMAP/PLANCK constraint as a function of $M_C$ for
  different values of $M_D$ : 300 GeV (dotted blue), 500 GeV (dashed green), and 1 TeV (solid red).
Right: constraints from the  FERMI and  HESS searches for monochromatic gamma--ray lines 
in the plane ($M_C$,$M_D$). (The area below the curve for a given $\kappa$ is excluded.)   }
  \label{Fig:relic}
\end{figure}

\noindent
Assuming that dark matter is thermally produced, its abundance should be consistent with
the WIMP freeze--out paradigm.  
As explained above, the only DM annihilation channel is $CC\rightarrow VV$ with $V=\gamma,Z$.
The corresponding cross section must be in a rather narrow window to fit observations.
The left panel of   Fig.~\ref{Fig:relic} shows parameter space consistent with the  
WMAP/PLANCK  measurements  \cite{Jarosik:2010iu,Ade:2013xsa}  
of the DM relic abundance for different values of 
$\kappa$, $M_C$ and $M_D$. For generality,  we allow for vastly different $M_C$
and $M_D$ in our numerical analysis. In the case $M_C^2 \ll M_D^2$, the scaling 
behaviour $\langle \sigma v \rangle \sim \kappa^4/M^2$  of Eq.~\ref{annih}   is replaced by
\begin{equation} 
\langle \sigma v \rangle \sim \kappa^4 \; {M_C^2 \over M_D^4 } \;,
\end{equation}
which stems from the momentum factors at the vertices.
Thus,  the annihilation cross section grows with the dark matter mass and, in turn, 
the WMAP/PLANCK--allowed $\kappa$'s decrease with increasing  $M_C$. 
The former take on rather natural values of order one for $M_D$
between 100 GeV and  several TeV.
The main annihilation channel is $CC \rightarrow \gamma\gamma$, which for 
$M_C \simeq M_D \simeq 200$ GeV constitutes about 60\% of the total cross section. 
The channels $CC \rightarrow \gamma Z$ and $CC \rightarrow Z Z$ contribute
35\% and 5\%, respectively. The allowed parameter space is subject to the 
FERMI and HESS constraints on the gamma--ray emission, which we study in the next
subsection.

\subsection{Indirect DM detection constraints}

\begin{figure}
    \centering
    \includegraphics[scale=0.6]{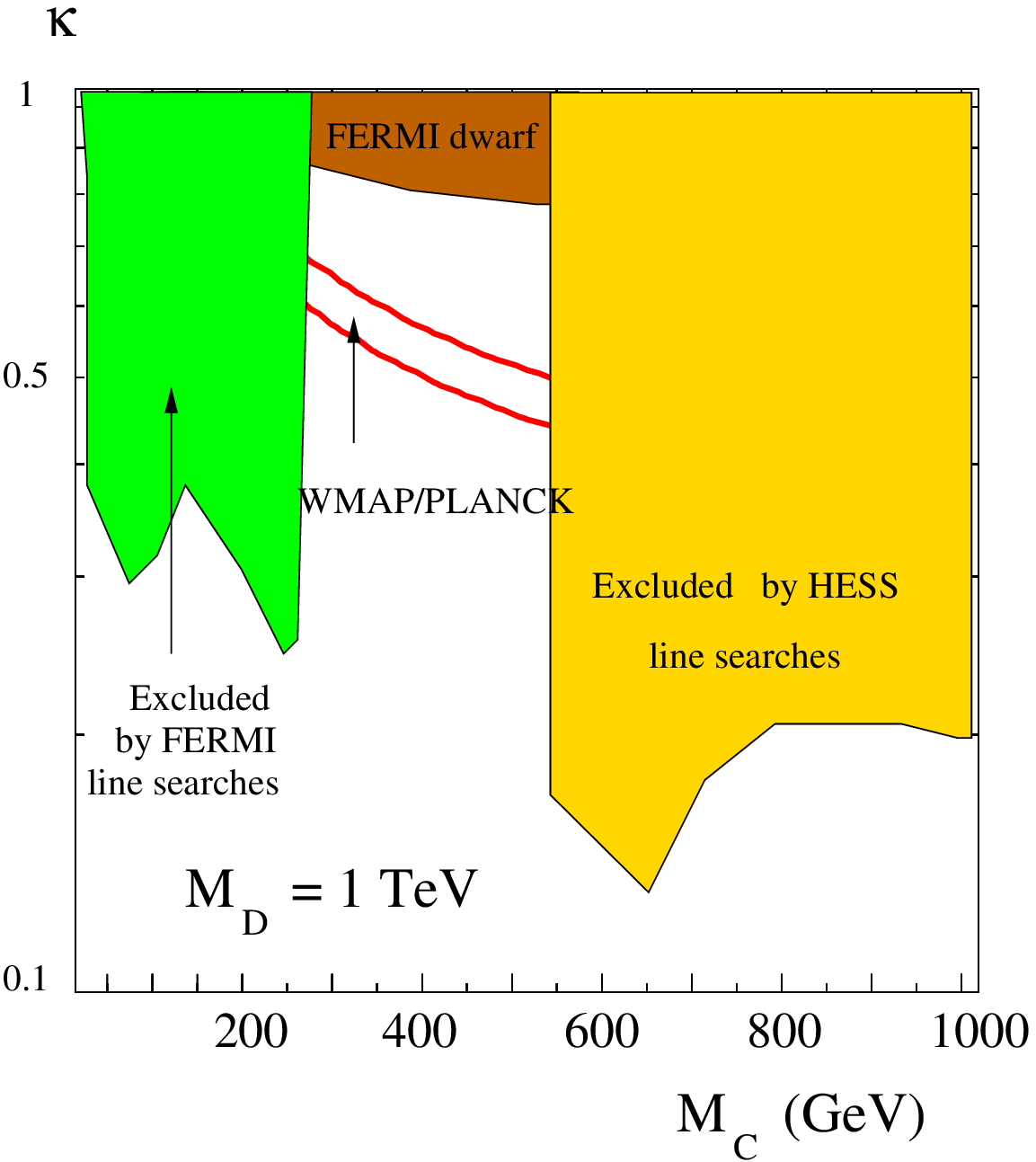}
    \hspace{0.5cm}
      \includegraphics[scale=0.6]{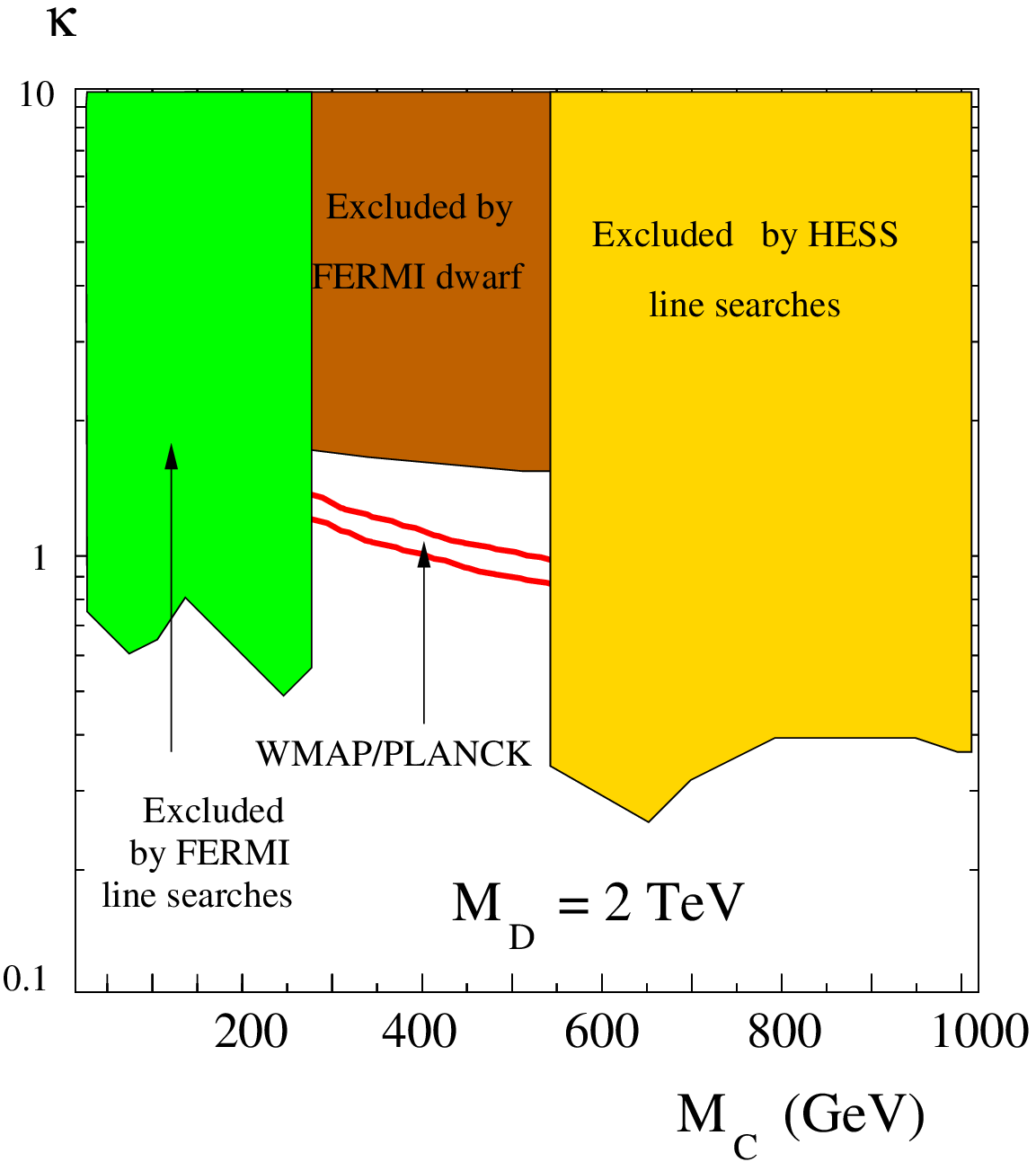}
  \caption{ FERMI and  HESS constraints on  gamma--ray monochromatic lines and continuum   
in the plane  ($M_C$, $\kappa$) 
  for $M_D =1$ TeV [left] and 2 TeV [right]. The area between the red lines is consistent
with thermal DM relic abundance.}
  \label{Fig:fermibis}
\end{figure}

Dark matter can be detected indirectly by observing products of its annihilation 
in regions with enhanced DM density.
The main feature of the Chern--Simons--type dark matter is that the dominant annihilation channel
leads to a di--photon  final state.  
These photons are   monochromatic due to the low DM velocity
nowadays ($v_C \simeq 300~\mathrm{km s^{-1}}$), which is a ``smoking--gun'' signature of 
our model.   The proportion of the di--photon final state increases somewhat compared
to that in the Early Universe due to the (slight) reduction of the center--of--mass energy
of the colliding DM particles. In particular,
for  $M_C \simeq M_D \simeq 200$ GeV,
the channels  $CC \rightarrow \gamma \gamma$, $CC \rightarrow \gamma Z$ and $CC \rightarrow Z Z$
constitute  approximately 63\%, 33\%, 4\% of the total cross section.  
One therefore expects  an intense monochromatic gamma--ray line at  $E_\gamma = M_C$ and a weaker
line at $E_\gamma = M_C - M_Z^2/(4 M_C)$. Such lines would provide convincing evidence for 
DM annihilation since astrophysical processes are very unlikely to generate such a photon
spectrum.

Recently, FERMI \cite{Ackermann:2012qk,Fermi-LAT:2013uma} and 
HESS \cite{Abramowski:2013ax} collaborations
have released their analyses of the monochromatic line searches around the Galactic Center.
Due to its limited energy sensitivity, the FERMI satellite sets a bound on 
the di--photon annihilation cross section  $\langle \sigma v \rangle_{\gamma \gamma}$ in the 
DM mass range $1~\mathrm{GeV} \lesssim M_C \lesssim 300$ GeV.   HESS, on the other hand,
is restrained by  its threshold limitations and provides bounds in the DM mass range 
 $500 \;{\rm GeV}  \lesssim M_C \lesssim 20$ TeV.\footnote{ HESS reports its results for  the  Einasto DM distribution 
profile, while  FERMI has extended its study to other profiles as well. 
To be conservative, we use  the FERMI limits for  the isothermal profile.} 
Combining the two analyses allows us to eliminate large portions of parameter space as shown in 
Fig.~\ref{Fig:relic} [right] and Fig.~\ref{Fig:fermibis}. We note that increasing the mediator mass $M_D$ 
has the same effect as decreasing the coupling $\kappa$.
The important conclusion is that 
 FERMI and HESS exclude the possibility of  thermal DM relic abundance in the relevant mass ranges.
Indeed, their bounds are of order $\langle \sigma v\rangle_{\gamma\gamma} \lesssim 10^{-27} \mathrm{cm^3 s^{-1}}$,
whereas thermal dark matter requires $\langle \sigma v \rangle \simeq 10^{-26} \mathrm{cm^3 s^{-1}}$.

To fill the gap between 300 and 500 GeV where the monochromatic signal is not constrained, one can use the diffuse 
gamma--ray flux. 
Indeed, even though the FERMI energy cuf--off is at 300 GeV, annihilation of heavy particles produces a continuum 
photon spectrum which can be detected by FERMI. 
In our case, the continuum  
comes from  the $ZZ$ and $Z\gamma$ final states with subsequent $Z$--decay. Since such final states contribute about 
40\% to the total cross section, the resulting constraint is not very strong. 
There exist several analyses of  bounds on DM annihilation in the  galactic halo \cite{Ackermann:2012rg}, galactic center  \cite{Hooper:2012sr} and   dwarf galaxies \cite{DrlicaWagner:2012ry}.
The latter provides the strongest FERMI constraint at the moment, while that from  HESS is very weak,   
and we use it to restrict  our parameter space 
(Fig.~\ref{Fig:fermibis}). The conclusion is that thermal DM  
in the 300--500 GeV mass range remains viable and can soon be tested by HESS/FERMI.

\begin{figure}
    \centering
    \includegraphics[scale=0.6]{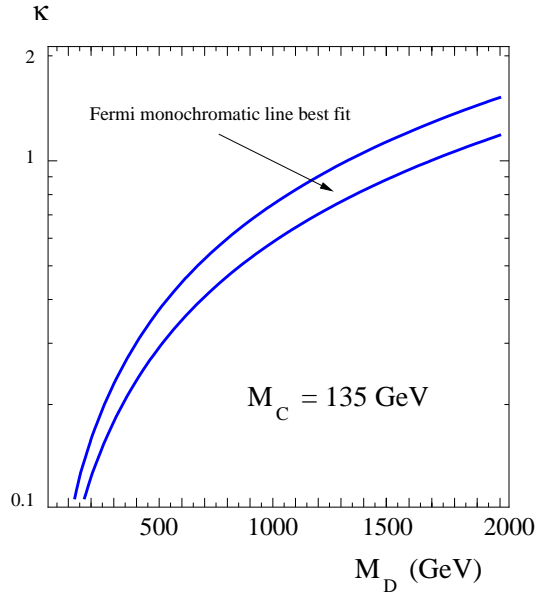}
  \caption{ Parameter space (between the lines) satisfying
  $\langle \sigma v \rangle_{\gamma \gamma} = (1.27 \pm 0.32^{+0.18}_{-0.28})\times 10^{-27} ~\mathrm{cm^3 s^{-1}}$ 
  and fitting the tentative FERMI gamma--ray line at 135 GeV.}
  \label{Fig:wenigerline}
\end{figure}

\subsection{On the tentative 135 GeV gamma--ray line}

When analyzing FERMI data, several groups  found some indications of a monochromatic (135 GeV)
gamma--ray line from the galactic center \cite{Weniger:2012tx,Raidal}. The significance of the ``signal'' appears to be around
3.3 sigma taking into account the look--elsewhere effect, although this has not been confirmed by 
the FERMI collaboration. A somewhat optimistic interpretation of the line is that it could be due to 
DM annihilation at the galactic center (see \cite{Chalons:2012hf,Profumo:2012tr,Ibarra:2012dw}
for recent discussions), with the cross section 
$\langle \sigma v \rangle_{\gamma \gamma} = (1.27 \pm 0.32^{+0.18}_{-0.28})\times 10^{-27} ~\mathrm{cm^3 s^{-1}}$ 
for an Einasto--like profile \cite{Weniger:2012tx}.

In this work, we will be impartial as to whether the line is really present in the data or not.
Instead, we use the analysis of \cite{Weniger:2012tx} as an example to show that the hypercharge
portal can easily accommodate a monochromatic signal from the sky.
Our result is shown in Fig.~\ref{Fig:wenigerline}. Having fixed $M_C=135$ GeV, we observe that 
the gamma--ray line can be accommodated for any mediator mass $M_D$. As explained above, the continuum
constraint is inefficient here since it applies to subdominant final states. On the other hand,
the required annihilation cross section is too small for  DM to  be a thermal relic.

\begin{figure}
    \centering
    \includegraphics[scale=0.6]{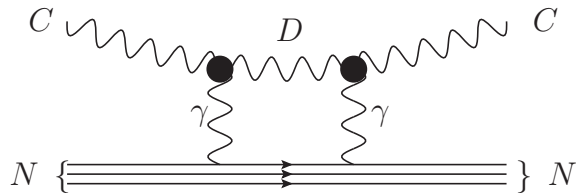}
  \caption{Dark matter scattering off a nucleon.}
  \label{Fig:feynman2}
\end{figure}

\subsection{Direct detection constraints }

An important constraint on properties of dark matter is set by direct detection experiments
which utilize possible DM interactions with nuclei.
In our case, 
dark matter scattering off nuclei is described by the 1--loop diagram of Fig.~\ref{Fig:feynman2}
together with its cross--version, and similar diagrams with $Z$--bosons in the loop. 
Setting for simplicity $M_C \simeq M_D =M$, 
we find that in the non--relativistic limit this process is described by the operators
\begin{eqnarray}
&& O_{SI} \sim {\alpha \kappa^2 \over 4 \pi} {m_N \over M^2} \; {\overline \Psi} \Psi \; C^\mu C_\mu \;, \nonumber\\
&& O_{SD} \sim {\alpha \kappa^2 \over 4 \pi} {1\over M^2} \;  \epsilon_{\mu \nu \rho \sigma}
{\overline \Psi} \gamma^\mu \gamma^5 \Psi \; C^\nu i\partial^\rho C^\sigma \;,
\end{eqnarray}
where $m_N$ is a hadronic scale of the order of the nucleon mass and $\Psi$ is the nucleon spinor.
$ O_{SI}$ and $O_{SD}  $ are responsible for spin--independent and spin--dependent scattering,
respectively. The former is suppressed both by the loop factor and the nucleon mass, while the latter
is suppressed by the loop factor only. The resulting cross sections are quite small,
$\sigma_{SI} \sim \kappa^4/M^2 \; (\alpha/4\pi)^2 (m_N / M)^4 \sim 10^{-46}{\rm cm^2}$ for 
$\kappa \sim 1$ and $M\sim 100$ GeV, whereas the spin--dependent cross--section is 
of the order of $\sigma_{SD} \sim \kappa^4/M^2 \; (\alpha/4\pi)^2 (m_N / M)^2 \sim 10^{-42}{\rm cm^2}$
for the same parameters. The current XENON100 bounds are   
$\sigma_{SI} \lsim {\cal O} (10^{-45})
{\rm cm^2}$ \cite{Aprile:2012nq} and $\sigma_{SD} \lsim {\cal O} (10^{-40}) {\rm cm^2}$
\cite{Aprile:2013doa} for the DM mass around 100 GeV (which maximizes the XENON100
sensitivity). We thus conclude that no significant bounds
on $\kappa$ can be obtained from direct detection experiments. 
Furthermore, since the gamma--ray constraints require $\kappa < {\cal O}(10^{-1})$ in this mass range,
the prospects for direct DM detection are rather bleak, orders of magnitude beyond the 
 projected sensitivity of XENON1T  \cite{Beltrame:2013bba}.

%%%%%%%%%%%%%%%%%%%%%%%%%%%%%%%%%%%%%%%%
\subsection{LHC monojet constraints }
%%%%%%%%%%%%%%%%%%%%%%%%%%%%%%%%%%%%%%%%

 \begin{figure}
    \centering
    \includegraphics[scale=0.8]{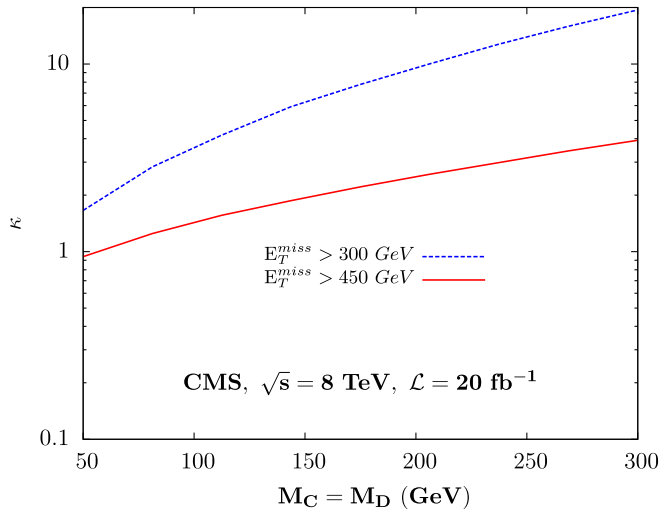}
  \caption{ Limit on $\kappa$ from monojet searches at CMS for $\sqrt{s}=$8 TeV and 
20 fb$^{-1}$ integrated luminosity.}
  \label{Fig:LHC}
\end{figure}

The vector states $C$ and $D$ can be produced at the LHC. If their mass difference is 
not sufficiently large, the photon coming from $D$--decay would not pass the 
experimental cut on the photon energy ($p_T>150$ GeV). In this case, production of
$C$ and $D$ would appear as missing energy. The latter can be detected in conjunction
with a jet coming from initial--state radiation, which sets a bound on DM production
(see also \cite{Djouadi:2012zc}).

In this subsection, we estimate the sensitivity of current monojet searches at the
LHC to dark matter production through its coupling to $Z$ and $\gamma$.
Our constraints are based 
on the search
for monojets  performed by the CMS collaboration  which makes use of
19.5 fb$^{-1}$ of data at 8 TeV  center of mass energy \cite{cms_mono}.     The
basic selection requirements used by the CMS experiment for monojet events 
are as follows:\vspace*{-4mm}   

\begin{itemize} 
\item at least 1 jet with $p_T^j > 110$ GeV and $|\eta^j| < 2.4$;\vspace*{-3mm} 

\item at most 2 jets with $p_T^j > 30$ GeV;\vspace*{-3mm}   

\item no isolated leptons.\vspace*{-3mm} 

%\item \mbox{missing transverse momentum $p_T^{\rm miss} \ge
%250,300,350,400,450,500,550$ GeV.}\vspace*{-4mm}   
\end{itemize}

The CMS collaboration quotes the event yields for 7 different cuts on the missing
transverse momentum $p_T^{\rm miss}$ between $250$ and $550$ GeV.  These are
largely dominated by the SM backgrounds, namely  $Z$+jets, where the $Z$ boson
decays invisibly, and $W$+jets, where the $W$ boson decays leptonically and the
charged lepton is not reconstructed. In particular, with 19.5 fb$^{-1}$ data, the
CMS collaboration estimates the background to be $18506 \pm 690 (1931 \pm 131)$
events for
$p_T^{\rm miss} >300 \;(450)$ GeV.     

A virtual $Z$--boson or a photon produced with a significant transverse
momentum and coupled to  invisible states  can 
also lead to the  topology that is targeted by the monojet searches.   
In order to estimate the sensitivity of the CMS monojet search to the 
``$Z/\gamma \rightarrow$ invisible''   signal, 
we generate the $pp \! \to \!Z/\gamma \!+ \!{\rm jets} \to {\rm
C D} \!+ \!{\rm jets}$ process at the parton level with 
 {\tt Madgraph 5} \cite{Alwall:2011uj}. 
Showering and hadronization is performed using {\tt Pythia
6}~\cite{Sjostrand:2006za}, while {\tt Delphes 1.9}~\cite{Ovyn:2009tx}  is employed
to simulate the ATLAS and  CMS detector response. We have imposed the analysis cuts listed above on
the simulated events  to find the signal efficiency.   
As a cross-check, we have passed $(Z \to \nu \nu)$ + jets background events through the
same simulation chain, 
obtaining efficiencies consistent with the data--driven estimates of that background
provided by CMS.

We use the total event cross section to put constraints on the dark matter coupling to
the $Z/\gamma$ gauge bosons.
We compute the observed 95\%CL exclusion limits on the dark matter--SM coupling
$\kappa$ for  given  masses ${M_{C},M_{D}}$ by requiring 
(see, e.g.~\cite{Fox:2011pm})
 \begin{equation}
 \chi^2 =
\frac{(N_{obs}-N_{SM}-N_{DM}(M_{C},M_{D},\kappa))^2}{N_{SM}+N_{DM}(M_{C},M_{D},\kappa)+\sigma_{SM}^{2}
} = 3.84 \;.
\end{equation}
Here $N_{obs}$ is the number of observed events, $N_{SM}$ the number of expected
events, $N_{DM}$ the number of expected signal events and $\sigma_{SM}$ being the
uncertainty in the predicted number of backgrounds events. 
The expected strongest bounds should come from the analysis with the hardest
$p_T^{\rm miss}>550$~GeV cuts, but the strongest observed bound come from
the $p_T^{\rm miss}>450$~GeV cuts due to an important downward fluctuations in the
data.
Fig.~\ref{Fig:LHC} shows the resulting limits on  $\kappa$  
 for two different sets of cuts, $p_T^{\rm miss}>300$~GeV and  $p_T^{\rm
miss}>450$~GeV, with the latter   providing the best limit. 
We see that the current monojet bounds are relatively weak, $\kappa < {\cal O}(1)$
for $M_C \sim M_D \sim 100$ GeV, and not competetive with the constraints from the
monochromatic gamma--ray searches.

%%%%%%%%%%%%%%%%%%%%%%%%%%%%%%%%%%%%%%%%
\subsection{LHC monophoton constraints }
%%%%%%%%%%%%%%%%%%%%%%%%%%%%%%%%%%%%%%%%

 \begin{figure}
    \centering
    \includegraphics[scale=0.8]{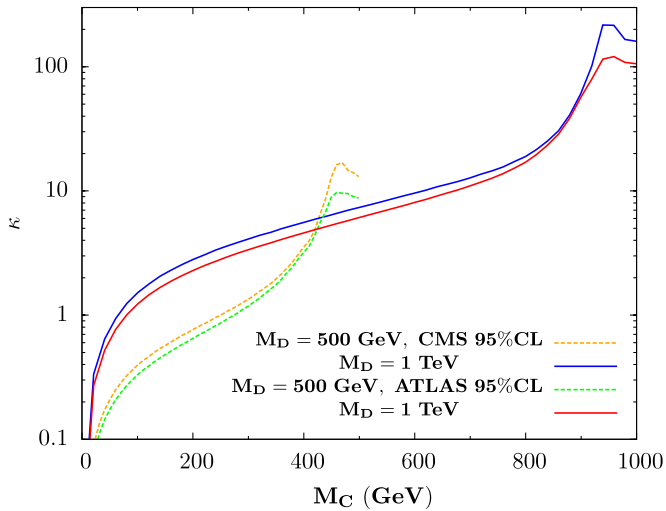}
  \caption{ Limits on $\kappa$ from monophoton searches at 
CMS (5 fb$^{-1}$ at $\sqrt{s}=$7 TeV)   and ATLAS 
(4.6 fb$^{-1}$ at $\sqrt{s}=$7 TeV).  }
  \label{Fig:LHC1}
\end{figure}

Another characteristic collider signature of vector DM production is monophoton emission plus
missing energy. In this case, $C$ and $D$ are produced on--shell through the photon or $Z$, while
their mass difference must be sufficiently large such that
$D$ decays inside the detector and the photon energy is above the threshold.  
We rely on the search
for a single photon  performed by the CMS collaboration  which makes use of
5 fb$^{-1}$ of data at 7 TeV  center of mass energy \cite{Chatrchyan:2012tea} and the one performed by
the ATLAS collaboration which makes use of
4.6 fb$^{-1}$ of data at 7 TeV  center of mass energy \cite{Aad:2012fw}.
The basic selection requirements used by the CMS experiment for monophoton events
are as follows:\vspace*{-4mm}

\begin{itemize}
\item 1 photon with $p_T^\gamma > 145$ GeV and $|\eta^\gamma| < 1.44$;\vspace*{-3mm}

\item  $p_T^{\rm miss} > 130$ GeV;\vspace*{-3mm}

\item no jet with $p_T^j > 20$ GeV that is $\Delta R > 0.04$ away from the photon
candidate;\vspace*{-3mm}

\item no jet with $p_T^j > 40$ GeV and $|\eta^j| < 3.0$ within $\Delta R < 0.5$ of the axis of the
photon;\vspace*{-3mm}
\end{itemize}

\noindent Analogous  requirements used by  ATLAS  
are:\vspace*{-4mm}

\begin{itemize}
\item 1 photon with $p_T^\gamma > 150$ GeV and $|\eta^\gamma| < 2.37$;\vspace*{-3mm}

\item  $p_T^{\rm miss} > 150$ GeV;\vspace*{-3mm}

\item no more than 1 jet with $p_T^j > 30$ GeV and $|\eta^j| < 4.5$;\vspace*{-3mm}

\item $\Delta \Phi(\gamma,p_T^\gamma) > 0.4$, $\Delta R(\gamma,jet) > 0.4$ and $\Delta \Phi(jet,p_T^{\rm
miss}) > 0.4$;\vspace*{-3mm}
\end{itemize}

\noindent
The event yields obtained by ATLAS and CMS are
largely dominated by the SM backgrounds, namely  $Z$+$\gamma$, where the $Z$ boson
decays invisibly, and $W$+$\gamma$, where the $W$ boson decays leptonically and the
charged lepton is not reconstructed. Since ATLAS accepts events with one jet, $W/Z$+jets is also an
important background for the ATLAS analysis.
With 4.6 fb$^{-1}$ data, the
ATLAS collaboration estimates the background to be $137 \pm 18 (stat.) \pm 9 (syst.)$ events and
observed $116$ events. The analogous numbers for CMS with 5 fb$^{-1}$ are 
$75.1 \pm 9.4$ and $73$ events, respectively.

In order to estimate the sensitivity of the ATLAS and CMS single photon search to DM production, 
we have generated the $pp \! \to \!Z/\gamma \! \to { C D} \! \to { C C} + \!{\rm
\gamma}$ process.  We have used the program {\tt Madgraph 5} \cite{Alwall:2011uj} for the channels at the
parton level.
Showering and hadronisation was performed using {\tt Pythia 6}~\cite{Sjostrand:2006za}  and {\tt Delphes
1.9}~\cite{Ovyn:2009tx}  was employed
to simulate the CMS detector response. We have imposed the analysis cuts listed above on the simulated events
 to find the signal efficiency and used the total event cross--section to constrain the DM coupling
to $\gamma$ and $Z$.
The observed 95\%CL exclusion limits on  $\kappa$ for  given
 ${M_{C},M_{D}}$ are obtained by requiring
\begin{equation}
 \chi^2 =
\frac{(N_{obs}-N_{SM}-N_{DM}(M_{C},M_{D},\kappa))^2}{N_{SM}+N_{DM}(M_{C},M_{D},\kappa)+\sigma_{SM}^{2} }
= 3.84 \;.
\end{equation}

The resulting limits on $\kappa$ for two choices of $M_{D}=500$~GeV and $M_{D}=1$~TeV
 are shown in Fig.~\ref{Fig:LHC1}. In the latter case, the bounds are relatively weak,
$\kappa < 1$ for $M_C >100$ GeV, and do not 
constrain the parameter space consistent with WMAP/PLANCK, FERMI and HESS (Fig.~\ref{Fig:fermibis}).
For $M_D=500$ GeV, the monophoton constraint is more important, although it  does not yet probe 
interesting regions of parameter space (Fig.~\ref{Fig:relic}).
In particular, it does not rule out 
the DM interpretation of the 135 GeV gamma--ray line (Fig.~\ref{Fig:wenigerline}). Indeed, for 
$M_C=135$ GeV, the LHC bound is about $\kappa < 0.5$, whereas the gamma--ray line requires $\kappa \sim 0.3$.

We thus find that the monophoton constraint is not yet competitive with the astrophysical/cosmological ones. 
We have also checked that no useful constraint is imposed by  searches for mono--$Z$ emission 
($D \rightarrow Z + C$), mostly due to its smaller production cross section.

\subsection{Summary of constraints}

For the DM mass above 100 GeV, the most relevant  laboratory
constraints are imposed by the LHC searches for monojets and monophotons.
The former are applicable for quasi--degenerate $C$ and $D$, while the latter
apply if there is a substantial mass difference between them. 
The monophoton constraint is rather tight for light DM, e.g. $\kappa <$ few$\times 
10^{-1}$ for $M_C \sim 100$ GeV and $M_D \sim 500$ GeV. 
This is stronger than the unitarity bound (\ref{uni}), which only applies 
for $\Lambda \gg M_{C,D}$. On the other hand, the monojet constraint is 
rather weak, $\kappa \lsim 1$.

The most important bounds on the model are imposed by astrophysical observations,
in particular, by FERMI and HESS searches for monochromatic gamma--ray lines. 
These exclude substantial regions of parameter space even for relatively
heavy dark matter, $M_{C,D} \sim 1$ TeV. Analogous bounds from continuum 
gamma--ray emission are significantly weaker as the latter is subleading in 
our framework (unlike in other models \cite{Buchmuller:2012rc}),
while direct DM detection is inefficient due to loop suppression.
These constraints still allow for thermal DM in the mass range 200--600 GeV
(Fig.~\ref{Fig:fermibis}). 

Finally, the model allows for an ``optimistic'' interpretation of the tentative
135 GeV gamma--ray line in the FERMI data. The line can be due to (non--thermal)  
dark matter annihilation with $M_C \simeq 135$ GeV for a range of the mediator
mass $M_D$. This interpretation is consistent  with  the constraints coming from  the 
continuum gamma--ray emission, direct DM detection  and the LHC searches.

\section{Conclusion}

We have considered the possibility that the hidden sector contains more than one
massive vector fields. In this case, an additional  $dim$--4 
interaction structure of the
Chern--Simons type becomes possible. It couples the hypercharge field strength 
to the antisymmetric combination of the massive vectors. 
If the latter are long--lived,
the phenomenological signatures of such a coupling include missing energy  in 
decays of various  mesons and $Z$, as well as monojet and monophoton production
at the LHC.

The hidden sector may possess a $Z_2$ symmetry, which would make the lighter vector field stable
and a good dark matter candidate. The characteristic signature of this scenario is 
monochromatic gamma--ray emission from the Galactic Center, while the corresponding continuum  
contribution is suppressed. We find that this possibility is consistent with other constraints,
including those from the LHC and  direct DM detection. Large portions of the allowed parameter
space can be probed both by indirect DM detection and the LHC monophoton searches.

{\bf Acknowledgements.} 
We are grateful to A. Ali for useful discussions.
This  work was partially supported by
the French ANR TAPDMS {\bf ANR-09-JCJC-0146}, the Spanish MICINNs
Consolider-Ingenio 2010 Programme  under grant  Multi- Dark {\bf
CSD2009-00064} and the Collaborative Research Center SFB 676 of the DFG,
``Particles, Strings, and the Early Universe''. 
Y.~M. and J.~Q.  also acknowledge  partial support from the European
Union FP7 ITN INVISIBLES (Marie Curie Actions, PITN- GA-2011- 289442) and
the ERC advanced grant Higgs@LHC.

\end{document}